\documentclass[twocolumn,prl,aps,superscriptaddress]{revtex4}
\usepackage{floatflt}
\usepackage{psfrag}
\usepackage{amsmath}   
\usepackage{amssymb}
\usepackage{amsfonts}
\usepackage{mathrsfs}
\usepackage{graphicx}
\usepackage[usenames]{color}
\usepackage[dvipsnames]{xcolor}
\usepackage[resetlabels,labeled]{multibib}
\newcites{S}{Suppl}

\begin{document}

\title{Magnetodispersion of Two-Dimensional Plasmon Polaritons}

\author{I.\,V.\,Andreev}
\author{V.\,M.\,Muravev}
\author{N.\,D.\,Semenov}
\affiliation{Institute of Solid State Physics RAS, Chernogolovka, 142432 Russia}
\author{A.\,A.\,Zabolotnykh}
\affiliation{Kotelnikov Institute of Radio-engineering and Electronics of the RAS, Mokhovaya 11-7, Moscow 125009, Russia}
\author{I.\,V.\,Kukushkin}
\affiliation{Institute of Solid State Physics RAS, Chernogolovka, 142432 Russia}

\date{\today}

\begin{abstract}
We have investigated the spectrum of two-dimensional (2D) plasmon polaritons over the full range of magnetic fields. In our study, we investigate a disk-shaped two-dimensional electron system (2DES) with a metallic gate on the backside of the substrate. Importantly, we show that 2D plasmon polaritons  hybridize with the TM$_0$ photonic mode of a dielectric waveguide formed by a sample substrate. We have developed a theory for plasmon-polaritons in an infinite 2DES. We find the experimental data to be in good agreement with the developed theory.
\end{abstract}


\maketitle

\section{I. INTRODUCTION}

The interaction of light and matter is at the heart of nearly every optical process found in nature~\cite{Gruner}. It has been established that the response of a two-dimensional electron system (2DES) to the incident electromagnetic radiation is dominated by plasmon excitations. Because of their compelling properties, two-dimensional (2D) plasmons have proven to be a versatile tool for integrating photonics with electronics. For example, 2D plasmons provide extreme subwavelength confinement of light while still offering a long lifetime. Also, plasmon velocity can be tuned over a wide range by changing the 2DES density or applying an external magnetic field. In this regard, screened plasmons in gated structures are especially promising as they allow for density tuning through the field-effect~\cite{Chaplik1972, Heitmann1984, Eisenstein2000, Muravev2007, Andress2012, Koppens2012, Basov2012, Aizin:13, Koppens2018, Bandurin2018}. 
{Gated plasmons have linear dispersion that has been studied over a broad range of wave vectors and in the absence of retardation effects~\cite{Muravev2007, Gubarev2015, Kallin1984, Pinczuk1988}.}

Charges in a two-dimensional layer are incapable of effectively screening a three-dimensional field of an incident electromagnetic wave. This fact leads to a strong hybridization of light and 2D plasma, which is expressed in the formation of a new quasiparticle~--- plasmon polariton~\cite{Kukushkin:03, Muravev:18}. Polaritonic effects are particularly strong when the quasi-static frequency of the 2D plasmon $\omega_{\rm plasmon}$ is comparable to the frequency of light $\omega_{\rm light}$ for a given wave vector $q$, and when retardation becomes significant. It was shown that retardation leads to a considerable reduction in the resonant plasma frequency~\cite{Kukushkin:03}. In particular, for practically important gated plasmons, the retarded frequency is given by~\cite{Chaplik2015, Andreev:21}:
\begin{equation}
    \omega_p = \dfrac{\omega_g}{\sqrt{1+\dfrac{V_p^2 \varepsilon}{c^2}}},
    \qquad \omega_g = \sqrt{\frac{n_s e^2 h}{\varepsilon \varepsilon_0 m^{\ast}}} q = V_p \, q,
    \label{renorm_p}  
\end{equation}
where $\omega_g$ and $V_p$ are the frequency and velocity of the gated plasmon in the quasi-static approximation. Here, $q$ is the wave vector of the plasmon, $h$ is the distance between the conducting gate and 2DES, $n_s$ and $m^{\ast}$ are the density and effective mass of the 2D electrons. $\varepsilon_0$ and $\varepsilon$ denote, respectively, the electric constant and effective permittivity of the semiconductor crystal between the back gate and 2DES. The parameter $V_p \sqrt{\varepsilon}/c$ is a quantitative measure of the strength of the retardation effects.

When a perpendicular magnetic field is applied, the plasmon resonance in a laterally confined 2DES splits into two modes --- the cyclotron (bulk) and edge magnetoplasmons~\cite{Allen1983, Mast1985, Glattli1985, Fetter1986, Volkov1988, Shikin1989}. However, retardation effects significantly modify the plasmon-polariton behavior in the magnetic field. Unfortunately, the fundamental magnetoplasma mode is very weak due to radiative damping, and quickly disappears against the background of plasmon harmonics in relatively low magnetic fields~\cite{Mikhailov:05}. This makes the experimental determination of magnetodispersion virtually impossible. Such crossover is manifested in a zigzag behavior of the observed bulk magnetoplasma modes~\cite{Kukushkin:03}.

In the present paper, we investigate the magnetodispersion of 2D plasmon polaritons over the full range of magnetic fields. It is accomplished employing a unique structure design with a metallic gate on its backside. The gate effectively suppresses the radiative damping of plasma waves, eliminating the problem of zigzag behavior. Experiments on disks with different diameters allow us to determine the spectra of plasmon-polariton waves in magnetic fields of up to $0.3$~T. Most importantly, these experiments reveal the mechanism responsible for coupling the incident light and 2D plasma. We establish that 2D plasmon polaritons hybridize with the TM$_0$ photonic mode of a dielectric waveguide formed by a substrate.

\section{II. EXPERIMENTAL TECHNIQUE}

The experiments were carried out on semiconductor samples with a single GaAs/AlGaAs quantum well. The samples had the 2D electron density $n_s=7.5\times10^{11}$~cm$^{-2}$ and the electron mobility $\mu=0.4\times10^6$~cm$^2/($V$\cdot$s$)$ at the temperature $T=4.2$~K, corresponding to the transport relaxation time $\tau=15$~ps. Samples with disk-shaped mesas of various diameters were fabricated from the quantum well through optical lithography and wet chemical etching. In our study, we investigated six samples with mesa diameters $d=3, 4, 5, 6, 7,$ and $10$~mm. GaAs substrate thickness for all the samples was fixed at $h=640$~$\mu$m. A metallic gate was thermally deposited on the opposite to the quantum well side of the substrate. The sample schematic is depicted at the top of Fig.~\ref{fig1}. We used two schemes to excite microwave resonances in the 2DES.  One approach was to place the sample in the cross-section of a rectangular waveguide powered by a microwave generator. The other approach involved feeding the microwave signal from the generator to a wire antenna in the vicinity of the sample. To detect the microwave absorption in the sample, we employed a non-invasive optical technique based on the high sensitivity of GaAs recombinant photoluminescence spectra to even slight heating of 2DES. A detailed description of the technique is provided elsewhere~\cite{Kukushkin2002}. All the experiments were performed at the temperature $T=4.2$~K in the magnetic field of ($0$--$0.4$)~T directed perpendicular to the sample surface. The samples were immersed in a liquid-helium cryostat with a superconducting coil.

\section{III. RESULTS AND DISCUSSION}

\subsection{A. Magnetoplasmons in screened disks}

\begin{figure}[!t]
\includegraphics[width=\linewidth]{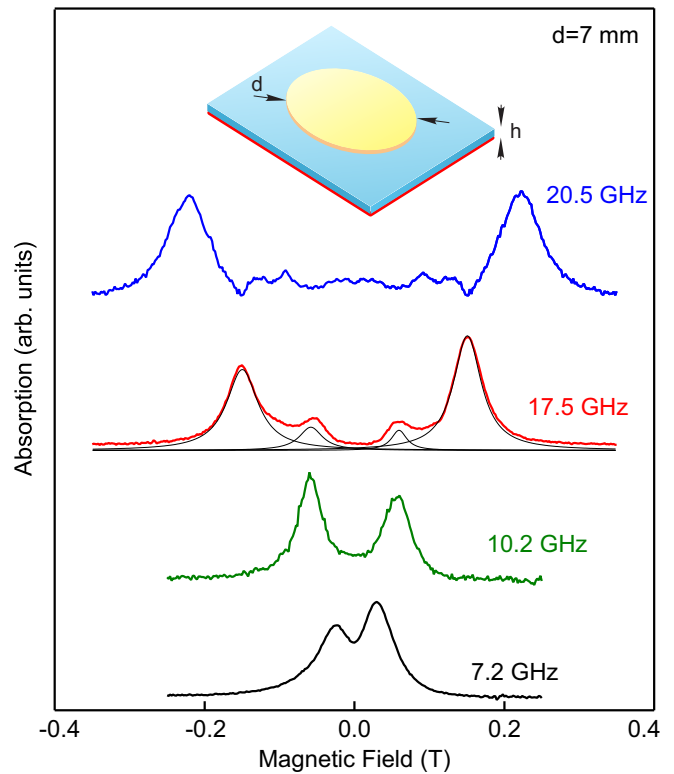}
\caption{Dependence of the microwave absorption on magnetic field in the sample with 2DES diameter $d=7$~mm, electron density $n_s=7.5\times10^{11}$~cm$^{-2}$, and substrate thickness $h=640$~$\mu$m. Plots show the data obtained at four excitation frequencies. Traces of Lorentzian fits to the experimental data for the frequency $f=17.5$~GHz are plotted with fine black curves. The inset at the top is the sample schematic with the back gate marked in red at the bottom surface of the substrate.}  
\label{fig1}
\end{figure} 

Figure~\ref{fig1} demonstrates typical microwave absorption dependencies on the magnetic field in a sample with disk diameter $d=7$~mm measured at different microwave excitation frequencies. The experimental curves reveal several microwave resonances symmetric with respect to magnetic field inversion. For accurate analysis, we fitted each experimental curve with several Lorentz peaks to determine the peak positions. An example of such a fit is shown in Fig.~\ref{fig1} for the data corresponding to the microwave excitation frequency $f=17.5$~GHz.

\begin{figure}[!t]
\includegraphics[width=\linewidth]{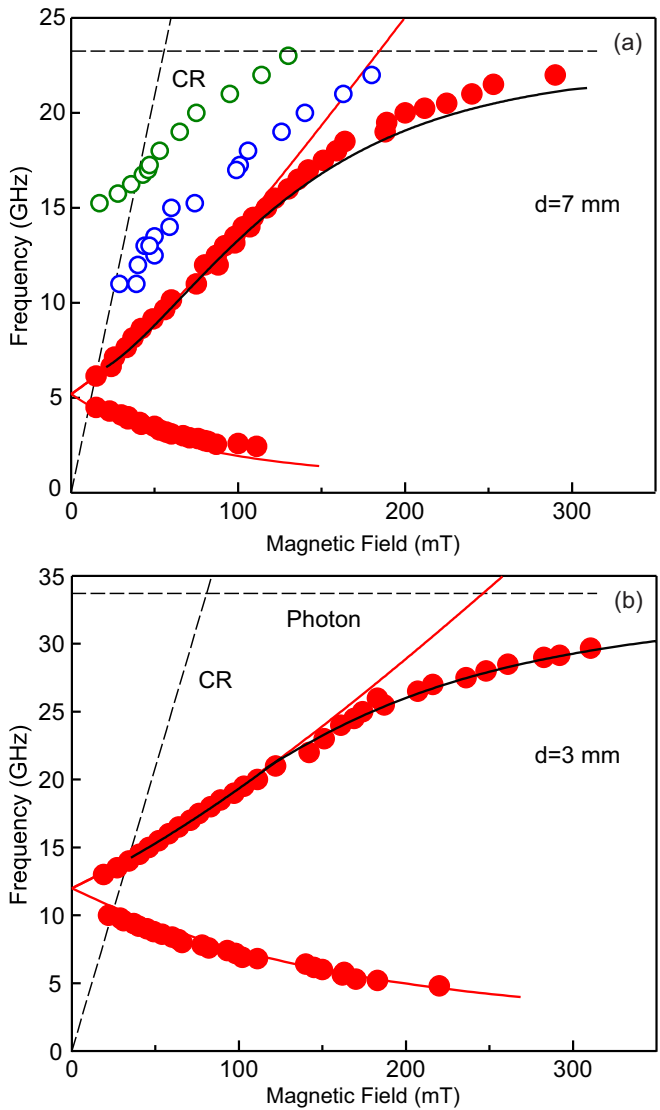}
\caption{Magnetodispersion of the plasmon modes observed in the samples with 2DES disk diameters $d=7$~mm (a) and $d=3$~mm (b). Both samples have the same electron density $n_s=7.5\times10^{11}$~cm$^{-2}$ and substrate thickness $h=640$~$\mu$m. Red dots denote the fundamental mode. Blue circles correspond to its second harmonic. Green circles correspond to the axisymmetric plasmon mode. The inclined dashed line indicates the cyclotron resonance (CR) magnetodispersion $\omega_c = eB/m^{\ast}$ with the effective electron mass $m^{\ast}=0.067m_0$. The red solid curves are the fits to the experimental points obtained according to Eq.~(\ref{bf_disk}). The theoretical prediction~\cite{Supplementary} for the magnetodispersion of the upper plasmon mode is plotted in a solid black line. The horizontal dashed lines indicate the photon asymptote $\omega = \omega_{\rm{inf}}$ at the infinite magnetic field (see details in text).
} 
\label{fig2}
\end{figure} 

To analyze the physical properties of the observed microwave resonances, we plot their frequencies versus the applied magnetic field. Magnetodispersion data for the samples with diameters $d=7$ and $3$~mm are presented in Fig.~\ref{fig2}(a) and \ref{fig2}(b), respectively. We can see that in all the samples, the behavior of the fundamental plasmon mode is similar in a qualitative sense. In the presence of the magnetic field, it splits into two branches. Such splitting is common for the plasma modes in a disk-shaped samples~\cite{Allen1983, Mast1985, Glattli1985, Fetter1986, Shikin1989}. It corresponds to removing the mode degeneracy due to the azimuthal number in a nonzero magnetic field. The lower mode $\omega_{-}(B)$ with negative magnetodispersion is often called the edge magnetoplasmon mode~\cite{Allen1983, Mast1985, Glattli1985, Volkov1988}. The upper mode $\omega_{+}(B)$ has positive magnetodispersion and corresponds to the usual (''bulk'') 2D magnetoplasmon~\cite{Kotthaus1977}. 

As for the gated 2DES under consideration, the edge and bulk magnetoplasmons as well as their splitting in a weak magnetic field have been previously investigated in~\cite{Andreev:21}. In the same work, it has been established that given strong retardation, magnetodispersion of the modes in a screened disk is well described by the following equation~\cite{Allen1983, Shikin1989, Khisameeva2020, Marquez2020}:
\begin{equation}
    \omega_{\pm}=\pm \frac{\omega_c'}{2} + \sqrt{\left( \frac{\omega_c'}{2} \right)^2 + \omega_p^2},
\label{bf_disk}
\end{equation}
where $\omega_p$ is the plasma frequency at zero magnetic field determined by Eq.~(\ref{renorm_p}) and $\omega_c'$ is the renormalized cyclotron frequency
\begin{equation}
    \omega_c' = \dfrac{eB/m^{\ast}}{1+\dfrac{V_p^2 \varepsilon}{c^2}}.
    \label{renorm_c}
\end{equation}
For the screened plasmon in a disk geometry, the wave vector $q$ in Eq.~(\ref{renorm_p}) should be taken as $q=3.7/d$~\cite{Fetter1986}. The experimental points are fitted according to Eq.~(\ref{bf_disk}) (solid red lines in Fig.~\ref{fig2}), with $\omega_p$ and $\omega_c'/B$ used as free parameters. For example, considering a sample with $d=7$~mm in zero magnetic field, we find the plasma frequency of the fundamental mode to be $f_p=5.2$~GHz. It can be seen that Eq.~(\ref{bf_disk}) accurately describes the lower $\omega_{-}(B)$ and upper $\omega_{+}(B)$ branches of magnetodispersion in low magnetic fields, up to several tens of mT. In the high magnetic field limit, when $eB/m^{\ast} \gg \omega_{+}(B)$, the measured data strongly deviate from~(\ref{bf_disk}). 

To account for such behavior, we develop a rigorous theory of the plasmon-polariton waves in an infinite gated 2DES~\cite{Supplementary}. The theory is based on Maxwell's equations with standard electrodynamical boundary conditions: vanishing of the tangent components of plasmon electric field on the metal surface, their continuity at 2DES plane, and discontinuity of their derivatives due to nonzero 2D current density in 2DES. To find the dispersion equation, we also use material equations (Ohm's law and the continuity equation for charge density current) with the Drude model for the conductivity tensor. As a result, the developed theory makes it possible to calculate the plasmon frequency $\omega_+$ for a given wave vector and magnetic field. To compare this theory with the experiment, we use a standard wave vector quantization for plasmons in a gated disk-shaped 2DES, $q=3.7/d$~\cite{Fetter1986}. Fig.~\ref{fig2} includes the theoretical results for $\omega_+(B)$ calculated at high magnetic fields for $\omega_c > \omega_+(B)$ (solid black curves). It shows that in this range of magnetic fields, the theory is in good agreement with the experimental data. 

We note that it was impossible to determine the magnetodispersion of plasmon-polariton modes experimentally for a long time due to the crossover between different plasmon harmonics~\cite{Mikhailov:05}. It occurs because the strong interaction of the fundamental plasmon with light causes a substantial radiative decay of the mode. The fundamental mode decays faster and releases its oscillator strength to the second harmonic~\cite{Mikhailov:05}, leading to the zigzag behavior of the observed resonance. Surprisingly, we do not see such a pattern in our case. It can be explained by the suppressed radiative decay of the plasma modes due to the back gate effect, considering strong screening $qh \ll 1$. For comparison, the Supplementary material~\cite{Supplementary} includes the magnetodispersion of the plasmon-polariton resonance obtained for a back-gated disk with $d=1$~mm. In that case, the condition for strong screening is not satisfied, and the zigzag behavior comes out. Thus, the absence of a zigzag pattern allowed us to correctly and accurately study the magnetodispersion of plasmon-polariton modes in the limit of high magnetic fields.

\subsection{B. Plasmon-polariton dispersion}

\begin{figure}[!t]
\includegraphics[width=\linewidth]{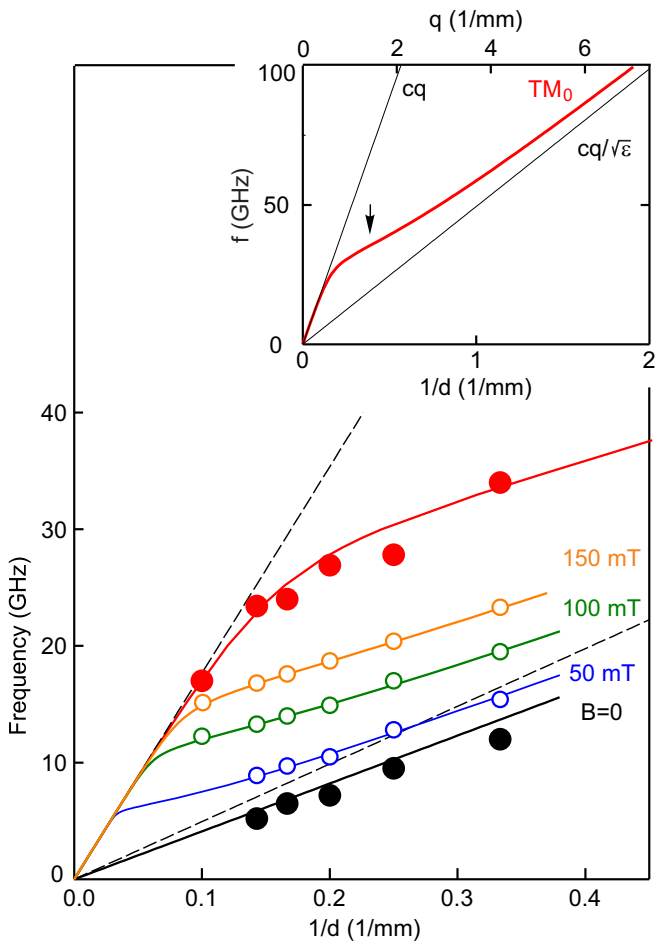}
\caption{Dispersion of 2D plasmon polaritons in various magnetic fields. Dashed lines correspond to the ``light cones" $cq$ and $cq/\sqrt{\varepsilon}$. Red dots denote the experimentally obtained asymptotics in high magnetic fields $\omega_{\rm{inf}}(q)$. The red curve is the dispersion of the TM$_0$ mode in the substrate. The black curve shows the prediction calculated from~(\ref{renorm_p}). The other solid curves are theoretical calculations described in the text. The inset displays the dispersion of the TM$_0$ dielectric waveguide mode propagating in the substrate (red curve). Arrow in the inset corresponds to the wave vector $q=1/h$.
}
\label{fig3}
\end{figure}

Identifying the exact mechanism of coupling between plasma waves and light has been a major challenge. To date, there is still no consensus on which photon mode a plasmon couples to. To address this critical question, we investigate the dispersion of plasmon-polariton waves at different values of the magnetic field $B$, from $B=0$ (black dots in Fig.~\ref{fig3}) up to the formal limit of the infinite magnetic field (red dots in Fig.~\ref{fig3}). At zero magnetic field, experimental data closely follow the theoretical prediction~(\ref{renorm_p}) (black line in Fig.~\ref{fig3})~\cite{Chaplik2015, Andreev:21}, given the retardation parameter $V_p \sqrt{\varepsilon}/ c=1.59$.  According to this theory, the plasma wave is hybridized with the TEM photon mode of the substrate, acting as a parallel-plate waveguide. The velocity of this TEM photon mode is $c/\sqrt{\varepsilon}$. 
{As the wave vector increases, plasmon-polariton dispersion is first transformed into that described in the case without retardation~\cite{Muravev2007, Gubarev2015} and then into the roton minimum~\cite{Kallin1984, Pinczuk1988}.}

At nonzero magnetic fields, the experimental data  shift upward in frequency (empty colored circles in Fig.~\ref{fig3}). 
{In relatively low magnetic fields, when $\omega_c' \ll \omega_p$, this shift equals to $\omega_c'/2$ (see Eq.~(\ref{bf_disk})). For example, at $B=50$~mT, the experimentally observed frequency shift is $\Delta f \approx 3.5$~GHz, which is in good agreement with $\omega_c'/4 \pi = 3$~GHz.} 
Dispersions for nonzero magnetic fields $B= 50, 100$,~and~$150$~mT predicted by the theory are plotted in colored curves in the same figure. The theory is in excellent agreement with the experimental data. For small wave vectors, $q$, all the theoretical dispersion curves converge to the light dispersion in vacuum $\omega = cq$ (dashed black line in Fig.~\ref{fig3}). Clearly, when $q$ is small, the wavelength $\lambda \gg h$, and the electromagnetic field of the plasmon-polariton wave is distributed in free space rather than in the substrate.

In the strong magnetic field limit, the $\omega_{+}(B)$ magnetodispersion curve flattens, tending asymptotically to a certain horizontal line $\omega = \omega_{\rm{inf}}(1/d)$ (the dashed horizontal lines in Fig.~\ref{fig2}). This behavior is typical of the hybridization of plasmons with photonic modes~\cite{Muravev:11}. To establish a particular photon mode that hybridizes with plasma excitations, we accurately determine $\omega_{\rm{inf}}(1/d)$ from the experimental data, based on the developed theory. For this purpose, we derive an asymptotic formula describing the behavior of the upper magnetodispersion branch $\omega_+(B)$ in the strong magnetic field limit $\omega_c \gg \omega_p$~\cite{Supplementary}:
\begin{equation}
	\label{Asym}
		\omega_+(B)=\omega_{\rm{inf}}-\frac{\rm const}{\omega_c^2-\omega_{\rm{inf}}^2} \approx \omega_{\rm{inf}}-\frac{\rm const}{\omega_c^2}.
\end{equation}
This formula is used to fit the curves to the experimental data. In Fig.~\ref{fig3}, the~$\omega_{\rm{inf}}$ frequencies are plotted as a function of the inverse disk diameter (red dots). The obtained dispersion starts from the free-space ``light cone" $cq$ at small wave vectors and then tends towards the ``light cone" in the substrate $cq/\sqrt{\varepsilon}$. Such behavior is typical of waveguide modes propagating in a dielectric slab.  For comparison, the red curve in Fig.~\ref{fig3} shows the dispersion of the TM$_0$ mode in a dielectric substrate with metal cladding on one side~\cite{Pozar}. The inset to Fig.~\ref{fig3} displays the TM$_0$ mode dispersion for a wide range of wave vectors. In the long-wavelength limit $qh < 1$, its frequency closely follows the free-space frequency of light $\omega = cq$. For the wave vector $q =3.7/d \approx 1/h$ (marked by an arrow in the inset of Fig.~\ref{fig3}), the mode dispersion trend changes direction and starts to asymptotically approach the $\omega = cq/\sqrt{\varepsilon}$.

\section{IV. CONCLUSION}

In conclusion, in the present paper, we have investigated the dispersion of plasmon-polaritons in disk-shaped 2DESs over the full range of magnetic fields. It has been accomplished by using back-gated samples, where the radiative decay of polariton modes is suppressed. Importantly, we have shown that the fundamental plasma mode couples to the TM$_0$ photonic mode of the dielectric substrate. This result is of critical importance to the integrated photonics, as it gives an understanding of which photonic modes need to be controlled to create new types of plasmonic devices.

\section{ACKNOWLEDGMENTS}

We thank V.A. Volkov for the stimulating discussions.
The authors gratefully acknowledge the financial support of the Russian Science Foundation (Grant No.~19-72-30003).

\newpage
\onecolumngrid
\appendix

\section{\large{Supplementary Material for\\ ``Magnetodispersion of Two-Dimensional Plasmon Polaritons''}} 

\subsection{I. MAGNETODISPERSION FOR THE DISK WITH $d=1$~MM}

\begin{figure}[!h]
\includegraphics[width=0.8\linewidth]{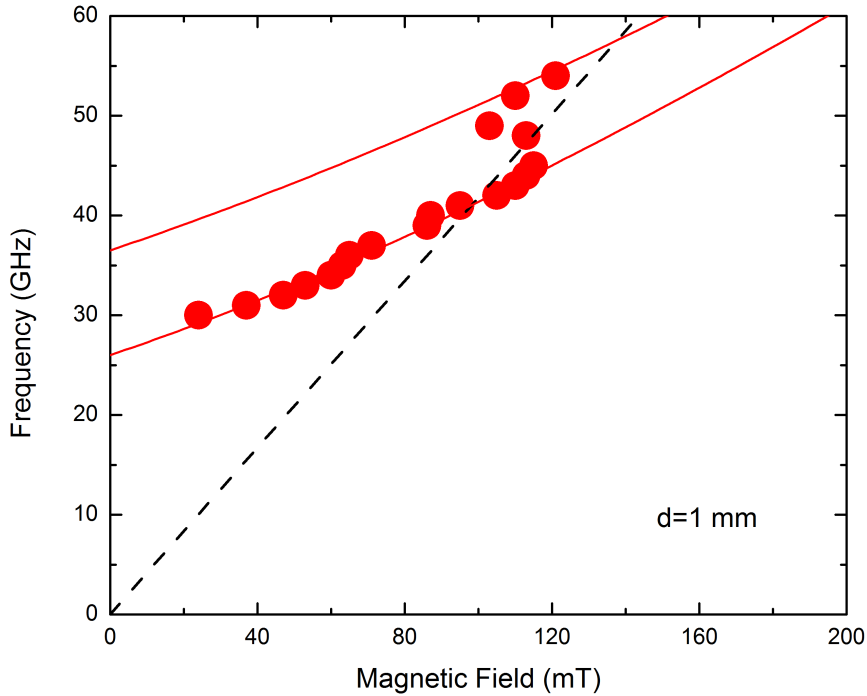}
\caption{Magnetodispersion of the fundamental plasmon-polariton mode (red dots) in a disk sample of diameter $d=1$~mm. The sample has the electron density $n_s=7.5\times10^{11}$~cm$^{-2}$ and the substrate thickness $h=640$~$\mu$m. The black dashed line indicates the cyclotron resonance (CR) magnetodispersion $\omega_c = eB/m^{\ast}$ with the effective electron mass $m^{\ast}=0.067m_0$. The red solid curves correspond to the magnetodispersions of the first two harmonics of the dimensional magnetoplasmon resonance.}   
\label{figS1}
\end{figure}

As stated in the main body of the paper, the magnetodispersion of the bulk (or cyclotron) magnetoplasmon resonances in laterally confined samples exhibits a complicated zigzag behavior~\cite{Kukushkin:03S, Mikhailov:05S}. Such a pattern is associated with the crossover between different plasmon harmonics. As the radiative decay of the fundamental mode occurs faster, it releases its oscillator strength to the next (second) harmonic~\cite{Mikhailov:05S}, and so on. We find the zigzag behavior to be more prominent in the case of the unscreened 2DES (see the main body of the paper).

Figure~\ref{figS1} illustrates the conception behind the zigzag behavior exhibited by the bulk magnetoplasmon modes. It shows the magnetodispersion plot (red dots) for the magnetoplasmon resonance measured in a back-gated 2DES disk sample of diameter $d=1$~mm and electron density $n_s=7.5\times10^{11}$~cm$^{-2}$. Given the substrate thickness $h=640$~$\mu$m, $h/D \sim 1$, which implies that the screening effect is weak. As indicated in the figure, the magnetodispersion of the observed resonance intersects the cyclotron magnetodispersion $\omega_c = eB/m^{\ast}$ with the effective electron mass $m^{\ast}=0.067m_0$ (black dashed line). Thus, it demonstrates the zigzag transition between the first and second harmonics of the dimensional magnetoplasmon resonance (lower and upper red solid curves in Fig.~\ref{figS1}, respectively).

\subsection{II. DISPERSION EQUATION FOR THE PLASMA WAVES IN A GATED 2DES}

Consider Maxwell's equations with standard electrodynamic boundary conditions (tangential components of the plasmon electric field vanishing on the metal surface, their continuity at the 2DES plane, and the discontinuity of their derivatives due to nonzero 2D current density in 2DES) together with material equations (Ohm's law and the continuity equation for the charge density current). Then, by analogy with Refs.~\cite{Kosevich1988, Jin2016}, we can find the dispersion equation for the plasma waves in an infinite 2DES with an ideal metal gate at a separation distance $h$. Considering a solution of the form $\exp(iqx-i\omega t)$, where $q$ is the 2D wavevector of the plasmon and the $x$-axis lies in the 2DES plane, the dispersion equation can be written as follows.
\begin{equation}
\label{disp_eq_rad}
	\left(\beta+\beta_\varepsilon \coth \beta_\varepsilon h- i\omega\mu_0\sigma_{xx}\right) \left(\frac{1}{\beta}+\frac{\varepsilon}{\beta_\varepsilon}\coth \beta_\varepsilon h+\frac{\sigma_{xx}}{-i\omega\varepsilon_0}\right)+\frac{\mu_0}{\varepsilon_0}\sigma_{xy}^2=0.
\end{equation}
Here, $\beta=\sqrt{q^2-\omega^2 /c^2}$, $\beta_\varepsilon=\sqrt{q^2-\omega^2\varepsilon/c^2}$, $\sigma_{xx}$ and $\sigma_{xy}$ are, respectively, the dynamic longitudinal and transverse (Hall) conductivities, $\varepsilon$ is the permittivity of the substrate between the 2DES and the gate, and $c$ is the speed of light in vacuum. The permittivity 'outside' the system is assumed to be unity.

Provided that the 2DES is 'clean', i.e. the electron relaxation time $\tau$ is large ($\omega\tau \gg 1$), in the framework of Drude model, 2D conductivities $\sigma_{xx}$ and $\sigma_{xy}$ can be expressed as:
\begin{equation}
\label{Drude}	
	\sigma_{xx}	=\frac{e^2n_s}{m^*}\,\frac{-i\omega}{-\omega^2+\omega_c^2},\quad \sigma_{xy}=\frac{e^2n_s}{m^*}\,\frac{-\omega_c}{-\omega^2+\omega_c^2},
\end{equation}
where $n_s$ is the electron concentration in the 2DES, $-e$ and $m^*$ are the electron charge and effective mass, and $\omega_c=eB/m^*$ is the electron cyclotron frequency in the 2DES.

Based on the Drude model, the dispersion equation can be formulated more explicitly as:
\begin{equation}
\label{disp_eq_rad2}
	\left(\left(\tilde{\beta}+\tilde{\beta_\varepsilon} \coth \tilde{\beta_\varepsilon}\right)(\omega_c^2-\omega^2)-A^2\omega^2\right) \left(\left(\frac{1}{\tilde{\beta}}+\frac{\varepsilon}{\tilde{\beta_\varepsilon}}\coth \tilde{\beta_\varepsilon}\right)(\omega_c^2-\omega^2)\frac{h^2}{c^2} +A^2\right)+A^4\omega_c^2=0,
\end{equation}
where $\tilde{\beta}=\beta h$, $\tilde{\beta}_\varepsilon=\beta_\varepsilon h$, and $A=n_s e^2 h/\varepsilon_0 m^{\ast} c^2$ is the dimensionless retardation parameter in a gated 2DES. Hence, given $A$ and $h$ parameters of the 2DES along with the magnetic field value expressed in terms of $\omega_c$, we can calculate the spectra of the bulk magnetoplasmon waves in a gated system.

\subsection{III. PLASMON POLARITON ASYMPTOTIC BEHAVIOR AT A HIGH MAGNETIC FIELD}

Considering a strong magnetic field, where $\omega_c \gg \omega$, and assuming that the Drude model is still applicable, we can obtain from Eq.~(\ref{disp_eq_rad2}) the asymptotic plasmon frequency for the upper magnetodispersion branch $\omega_+(B)$ as a function of the cyclotron frequency $\omega_c$:
\begin{equation}
	\label{Asym1}
		\omega_+(B)=\omega_{\rm{inf}}-\frac{const}{\omega_c^2-\omega_{\rm{inf}}^2} \approx \omega_{\rm{inf}}-\frac{const}{\omega_c^2}.
\end{equation}
Therefore, at low values of $1/B$, the dependence of $\omega_+$ on $1/B^2$ is linear, intersecting the $\omega$-axis at the point $\omega_{\rm{inf}}$. This result can be readily used to find the asymptotic frequency of a plasmon polariton in an infinite magnetic field $\omega_{\rm{inf}}$.

\end{document}